\documentclass[3p,times]{elsarticle}

\usepackage{algorithm}
\usepackage{multicol}
\usepackage{algorithmic, amsmath}
\usepackage{epsfig,graphicx}
\usepackage{enumerate}
\usepackage{calligra}
\usepackage{mathptmx}       
\usepackage{helvet}         
\usepackage{courier}        
\usepackage{type1cm}        
\usepackage{makeidx}         
\usepackage{graphicx}        
\usepackage{multicol}        
\usepackage[bottom]{footmisc}

\usepackage[figuresright]{rotating}

\begin{document}

\begin{frontmatter}




\title{Modeling and Performance Studies of Data Communication Networks using Dynamic Complex Networks}


\author{Suchi Kumari}
\address{Department of Computer Science and Engineering,\\
National Institute of Technology,\\
Delhi, India\\ email: {suchisingh@nitdelhi.ac.in}}
\author{Anurag Singh}
\address{Department of Computer Science and Engineering,\\
National Institute of Technology,\\
Delhi, India\\email: {Corresponding author: anuragsg@nitdelhi.ac.in}}

\begin{abstract}
All the existing real world networks are evolving, hence, study of traffic dynamics in these enlarged networks is a challenging task. The critical issue is to optimize the network structure to improve network capacity and avoid traffic congestion. We are interested in taking user's routes such that it is least congested with optimal network capacity. Network capacity may be improved either by optimizing network topology or enhancing in routing approach. In this context, we propose and design a model of the time varying data communication networks (TVCN) based on the dynamics of in-flowing links. Newly appeared node prefers to attach with most influential node present in the network. In this paper, influence is termed as \textit{reputation} and is applied for computing overall congestion at any node. User path with least betweenness centrality and most reputation is preferred for routing. Kelly's optimization formulation for a rate allocation problem is used for obtaining optimal rates of distinct users at different time instants and it is found that the user's path with lowest betweenness centrality and highest reputation will always give maximum rate at stable point.
\end{abstract}

\begin{keyword}

Data Communication Networks, Time varying Networks model, System Utility, Routing 
\end{keyword}

\end{frontmatter}

\section{Introduction}
Many real world networks, such as data communication network, World Wide Web, transportation network etc., follow heavy tailed power law degree distribution results in scale free behavior \cite{tadic}. These networks are evolving, hence, dynamics of the networks have been studied in past few years. In communication networks, data packets are transported from some source ($ S $) to destination ($ D $) via a path. The critical issue concerned with the communication networks is to optimize the network structure to improve searching or routing efficiency and avoid traffic congestion. The next issue is related to efficient usage of the capacity of the networks. It is observed that few nodes become congested and then, later it results in propagation of congestion in the whole network. It happens because these nodes are used most in the communication path as they are more central. Network capacity may be improved using two methods: (i) Methods to optimize network topology, and (ii) Enhancement in routing approach.
\subsection{Optimization of network topology }
Optimization of network topology corresponds to reorganization of the network for minimizing congestion and getting improved network capacity. Zhao \textit{et al.} \cite{onset} have enhanced network capacity by redistributing the load of heavily loaded nodes to other nodes in the network. Links are sorted in descending order according to their end node degrees and links are removed  in the descending sorted list \cite{liu}. For heterogeneous networks, Zhang proposed a method in which nodes having high betweenness centrality are removed first \cite{zhang}. Networks are growing during the whole of the life span and nodes are connected via preferential attachment (BA model) \cite{bara}. In BA model, appearance of new connections totally depends on the addition of new nodes to the system. But, in real systems, links are added and updated continuously. Tadic \cite{tadic} has developed a model for dynamics of directed networks. Degree distribution of both the outgoing and incoming links follow power law degree distribution and there is a correlation between them. It suggests that local structure of the network is different in comparison with the case of without updates. There is a need to include time parameter with the nodes and links of the basic network structure so, many models have been proposed for these time-varying graphs (TVGs) \cite{TVG,TVGdynamic,TME,smallTVG}. Model developed by Kostakos \textit{et al.} \cite{STE} is based on the concept of quick transfer and delayed transaction of data between distinct nodes in the communication network. Kim \textit{et al.} \cite{TME} focus their work on discovering closest and most betweenness central node  in the dynamically changing networks. A series of static graphs (i.e., the snapshots) are used to represent the network at a given time instant \cite{snapshot}. The snapshot model makes an external assumption that the network must be capable of retaining the link transitivity during the period of disconnection. Recently, Wehmuth \textit{et al.} \cite{TVG} have proposed a new unifying model for representing finite discrete TVGs. TVG nodes are considered as independent entities at each time and it is analyzed the proposed TVG is isomorphic to a static graph. It is an important result because of the use of the isomorphic directed graph as a tool to analyze both the properties of a TVG and the behavior of dynamic processes over a TVG. A framework to represent mobile networks dynamically in a Spatio-temporal fashion is designed and algebraic structural representation is also provided \cite{kumari2016}. 
\subsection{Enhancement in routing approach}
Network restructuring cost is more, hence most of the researchers focus on finding better routing strategies. Traditionally, packets are sent along the shortest paths in the term of hop count or minimal sums of weighted links \cite{SP}. Most of the shortest paths are passing through most central nodes and hence, sending data only through shortest path is not efficient. Yan \textit{et al.} \cite{efficient} have focused their work on searching efficient path, instead of searching of shortest paths. In communication networks, two rates are associated with each node: packet generation rate $ \lambda $ and packet forwarding rate $  \mu $. There is a critical rate  of packet generation $ (\lambda_c) $, below which network will be free from congestion. Two models are proposed by Zhao \textit{et al.} \cite{onset} for finding packet delivery rate of a node, (i) based on node's links and (ii) number of the shortest path passing through the node. In later model, critical rate is independent of network's size and topology, unlike, previous model. Du \textit{et al.} \cite{du} have developed a model in which packets are  generated with a rate $ \lambda $ between random source and destination pairs. Packets are propagated into the network such that load is minimized and will be delivered by Shortest Remaining Packet First (SRPF) algorithm. They also studied the behavior of congested node with the arrival of large number of packets. A heuristic routing strategies are used by Jiang \textit{et al.} \cite{jiang}, in which routing process is divided into $ N $ (size of network) steps. All the routes from the source node are based on the betweenness centrality and node's degree information. Pu \textit{et al.} \cite{pu} have formed a network model based on Barabasi Albert (BA) model and routing between pair of nodes are done such that cost function is minimized. Robustness and network efficiency are calculated against cascading failure. Networks are changing rapidly with time, hence, acquiring global information of the network  using global routing strategy is not useful for longer time. Various local routing strategies are proposed by Lin \textit{et al.} \cite{lin2016}. Proposed methods include node duplication avoidance, searching next nearest neighbors and restricted queue length. All the proposed strategies are used to enhance the performance of the system, by increasing packet generation rate of the nodes and reducing packet transmission time.
\subsection{Rate control mechanism}
Some nodes as well as links are getting added or removed but there is a net increment in the size of the networks with time. As network is expanding day by day so design and control of such network is a challenging work. Structure of networks is dependent on distribution of links  and directly affects the accessibility of particular node. Topology is changing and due to change in topology, the performance of the network is also affected. Communication network can respond to the randomly changing traffic but modern communication networks face multiple challenges at different layers. Real life communication networks are extremely volatile and making connections in the volatile environment with optimal network utility is a challenging task. Modeling their rate control behavior \cite{kelly2001} with volatile and dynamic connectivity helps us in getting optimal utilization of the system. La \textit{et al.} \cite{la2002} have extended Kelly's work by introducing a suitable pricing scheme. An algorithm is proposed for adjusting network's price and users' window size is re-framed to achieve optimal system utility by maintaining weighted proportionally fairness. Priya Ranjan \textit{et al.} \cite{ranjan2006} have investigated Kelly's optimization framework and used for finding stability conditions with the consideration of arbitrary delays. Effect of the non responsive traffic on the system stability is studied. It is shown that ubiety of such elements help in improving the stability of the system. Recently, progress is made on the earlier work of modeling of TVCN and a new Lyapunov function is being formulated for rate control scheme and long term stability is obtained \cite{pr2016}.

In previous researches, various routing strategies are proposed to avoid congestion in the communication networks. Here, work is focused on congestion and network utility both and the routing of the users are established with minimum congestion and maximum network utility along them. Network utility is calculated by summing up all individual user's utilities, and it will be maximum when, each user send data to their destinations with optimal rate. Once network is designed then, users select their routes for data communication with their destined nodes. In this paper, network is designed with the consideration of expansion, rewiring and removal of the links. A model is proposed for removing infrequently used links in the network apart from addition and rewiring of links. Addition and rewiring of links are based on preferential attachment. The removed links increase unnecessarily maintenance cost, hence, the removal of these links will reduce the cost. In the proposed model, the links attached to a node having lower degree, will be preferred to remove. Optimality of the routes are checked by analyzing Kelly's optimization approach \cite{kelly2001} for a rate allocation problem in communication networks at different time instants.

Centrality plays an important role for congestion in the network, especially betweenness centrality. Betweenness centrality of a node $ v $ is equal to the number of shortest paths from all node pairs pass through that node $ v $ and is given by,
\begin{equation*}
g(v) = \sum_{S \neq v \neq D} \frac{\sigma_{S \rightarrow D}(v)}{\sigma_{S \rightarrow D}}.
\end{equation*}
If larger number of shortest paths are passing through a node then, betweenness centrality and degree of that node will be more but, vice-versa is not true \cite{goh}. Connectivity distribution of both betweenness and degree centrality of the real world networks follow power law distribution with different but, correlated exponents \cite{betweenness}. Nodes with maximum betweenness centrality is the most congested node in the network \cite{onset}. In the proposed routing strategy, shortest paths with lowest total sum of betweenness centrality of nodes are selected for sending data packets. 

Some nodes are endorsed by the congested nodes and cause increase in load at that node. Hence, these nodes may be avoided for reducing data traffic in the network. Eigen vector centrality is used to check influence of one node to others. In this paper, influence is measured in terms of congestion in the networks. A node with high Eigen vector centrality is highly congested and it is undesirable to connect with that node. We have used a term, \textbf{``reputation"}, for finding overall congestion at any node. Reputation of a node is inversely related to Eigen vector centrality of the particular node, hence, most Eigen vector central node is least reputed. Both, betweenness centrality and reputation are collectively used for avoiding congestion in the network. Among all the shortest paths, the path having lowest betweenness central and most reputed are desirable for data communication between user pairs. These desirable paths are taken as an input for Kelly's optimization framework and are the achievers of optimal data rates $ x^* $.

Section 2 states background information about the related work and describe the mathematical model used in the analysis of rate control behavior of the user route. Section 3 introduces model for growth dynamics of TVCN, formulation of scaling exponent and the proposed routing strategies for communication network. Section 4 presents theoretical and simulation results, and in Section 5, conclusion and future research are discussed.
\section{Background and Related work}
In communication network, each node can generate and forward packets with distinct rate, $ \lambda $ and $ \mu $ respectively. There are three possible relationships between $ \lambda $ and $ \mu $: (i.) $ \lambda < \mu $ implies that system in free flow state, (ii.) $ \lambda = \mu $ shows boundary case for congestion and (iii.) $ \lambda > \mu $ allows system in congestion. Zhao \textit{et al.} \cite{onset} have proposed a model in which packet forwarding rate $ \mu $ of a node $ i $ for model: (i) based on degree of a node, $ \mu_i = 1 + \lfloor \beta k_i \rfloor $ and (ii) based on betweenness centrality of a node, $ \mu_i = 1 + \lfloor \frac{\beta g_i}{N} \rfloor $. Here, $ k_i $ = degree of node $ i $, $ g_i $ = betweenness centrality of node $ i $, $ N= $ size of the network and $ \beta $, $ 0 < \beta < 1 $. By theoretical estimation, the value of critical rate is given by $ \lambda_c = \frac{\mu_{Lmax} (N-1)}{g_{Lmax}} $. The node with maximum packet forwarding capacity and betweenness centrality are denoted by $ \mu_{Lmax} $ and $ g_{Lmax} $ respectively.\\
An order parameter $ \zeta(\lambda) $ \cite{rateeqn}, is used to describe the traffic and given by,
\begin{equation*}
 \zeta(\lambda) =  lim_{t \rightarrow \infty} \frac{\mu}{\lambda}\frac{\langle \Delta Num_p  \rangle}{ \Delta t}
\end{equation*} 
$ \Delta Num_p = Num_p(t + \Delta t)-Num_p(t) $, $ Num_p(t) $ is number of packets at time $ t $ and $ \langle . \rangle $ shows that average value over the time window $ \Delta t $. $ \langle \Delta Num_p  \rangle = 0 $ shows that no packet in the network and when $ \eta = 0 $, system is in free flow state. 
Yan \textit{et al.} \cite{efficient} have found average betweenness centrality $ g(k) $ for the nodes with same degree $ k $:
\begin{equation*}
g(k) = \frac{1}{N(k)} \sum_{v: k_v = k} g(v)
\end{equation*}
where, $ N(k) $ is number of nodes with degree $ k $. For Scale Free networks, $ g(k) \sim k^\alpha $. An efficient path $ P(i \rightarrow j) $ between node $ i $ and $ j $ is  calculated such that the sum $ L(P(i \rightarrow j:\alpha) = \sum_{i=0}^{N-1} k(x_i)^\alpha $ should be minimized for an optimal value of $ \alpha $. Critical rate, $ \lambda_c $ also depends on $ \mu $ and hence, $ \lambda_c(\beta) = \frac{N(N-1)}{g_{Lmax}^\beta} $. Liu \textit{et al.} \cite{liu} have studied network traffic by considering local and global routing strategies. Betweenness Centrality (globally calculated) of a link whose end nodes are $ m $ with degree $ k_m $ and $ n $ with degree $ k_n $, is linearly correlated with multiplication of degree of end nodes, $ k_m k_n $ (locally evaluated).  To avoid congestion in the network, links having maximum $ k_m k_n $ values are closed. After closing few links, $ \lambda_c $, average path length $ (L_{avg})  $ and average travel time $ (\langle T \rangle) $ are studied. Jiang \textit{et al.} \cite{jiang} have found heuristic method for routing and used an incremental approach for adding new paths in the network. As addition of new paths cause change in betweenness and degree of each node.
\begin{equation*}
L^s(P(i \rightarrow j)) = \sum_{v = 0}^{v=N-1} f^S(n_v)
\end{equation*}
where, $ f^S(n_v) = (g_{n_v}^S Deg_{n_v}^S)^\alpha $ is dynamic weight of node $ v $, $ n_v $ for source node $ S $, $ g_{x_v}^S $ is dynamic betweenness centrality of $ n_v $ and $ Deg_{n_v}^S $ is dynamic degree of node $ v $. Hence, weight of each link $ (i,j) $ is $ (g_i^SDeg_i^S)^\alpha+(g_j^S Deg_j^S)^\alpha $. All the previous works discussed till now are focused on routing and optimizing network structure. Tadic \cite{tadic} has discussed about restructuring of directed networks by considering growth and rearrangement at unique time scale. At each time unit $ t $ a new node is added to the network (growth) and a number $ X(t) $ and fraction $ \beta $ are used to find newly added and rearranged links in the network. (i.) A portion $ f_{add}(t) = \beta X(t) $ of new links are out-flowing links from the new added node $ i = t $ and (ii.) the remaining part $ f_{update}(t) = (1-\beta) X(t) $ are the updated links at other nodes in the network. 
Optimization of network topology and developing efficient routing strategies help us to avoid congestion and to improve overall network performance. In communication network, various users want to establish connection between specious node pairs via an optimal route. Routes are selected such that user can send data with optimal rate to their destinations. Kelly \cite{kelly2001} has formulated the rate allocation problem into optimization problem for a static network. We have used the formulation of rate control algorithm and proposed a mathematical model for dynamic communication networks. 
\subsection{Mathematical model}
A brief description is presented about mathematical representation of the dynamic networks. Network consists of $ N $ nodes and $ E $ links and life span of the network is $ T $. A set of $ R $ users are willing to send data in the network. Any link $ e_{mn} $ connects node $ m $ with node $ n $ can send maximum $ C_{e_{mn}} $ units of data through it, where, $ C_{e_{mn}} $ is capacity of the link $ e_{mn} $ and $ e_{mn} \in E $. In TVCN, each user $ r $ is assigned a route $ r \in R $ for a particular time instant $t_i \in T $. At the end of $ t_i^{th} $ time, a zero-one matrix $ A $ of the size $( N \times N )_{t_i} $ is defined where, $ A_{n_i, n_j, t_i} = 1 $, if node $ n_i $ and $ n_j $ are connected at time $ t_i $ otherwise zero. When the user $ r $ is assigned a rate $ x_{r, t_i} $ then, utility of user $r$ at rate $x_{r,t_i}$ is given as $U_{r, t_i}(x_{r, t_i})$ is
increasing, strictly concave function of $x_{r, t_i}$ over the range $x_{r,t_i} \geq 0$. Aggregate utility is calculated by summing up all
utilities of user $r$ with rate  $x_{r, t_i}$ and is denoted as $\sum_{r \in R, t_i} U_{r, t_i}(x_{r, t_i})$.
Rate allocation problem can be formulated as the following optimization problem.

\begin{eqnarray} 
&& SYSTEM(U_{t_i},A_{t_i}) \nonumber \\
&& maximize \sum_{r \in R, t_i} U_{r, t_i}(x_{r, t_i}) \label{e1} \\
&& A_{t_i} ^ Tx_{t_i} \leq C_{t_i} \mbox{ and }  x_{t_i} \geq 0  \nonumber
\end{eqnarray}

$A_{t_i}$ is a matrix at time interval $t_{i-1}$ to $t_i$.  The given constraint states that a link can not send data more than its capacity \cite{kelly2001}. It is difficult and unmanageable to allocate a suitable utility function and optimal rate to distinct users in complex networks. Hence, Kelly has divided this problem into two simpler problems named as user's optimal problem and network's optimal problem \cite{kelly2001}. Let user $ r $ decide an amount $ \mathcal{P}_r(t_i) $ to pay per unit time, while $ r $ is demanded a price $ \eta_r $  for unit flow. Hence, user will acquire a flow, $ x_r(t_i)= \mathcal{P}_r(t_i)/\eta_r$ at time $ t_i $. User's optimal price for sending $ x_r(t_i) $ at time $ t_i $ can be obtained using the optimization problem as 
\begin{eqnarray} 
\nonumber
&& User_r(U_r(t_i),\eta_r(t_i)),\nonumber \\
&& \mbox{ maximize } U_r(x_r(t_i))- \mathcal{P}_r(t_i), \label{e2} \\ 
&&  \mathcal{P}_r > 0 \nonumber
\end{eqnarray}
On the other hand, network wants to maximize weighted log function of $\mathcal{P}_r(t_i)$. Therefore, network optimization problem can be formulated as
\begin{eqnarray} 
\nonumber
&&NETWORK(A_{t_i},p_{t_i}),\\
&&maximize \sum_{r \in R, t_i} \mathcal{P}_r(t_i) log(x_r(t_i)), \label{e3}\\ \nonumber
&&A_{t_i}^T x_{t_i} \leq C_{t_i} \mbox{ and } x_{t_i} \geq 0. 
\end{eqnarray}

As the network is time varying so, the values of $ \eta_r $ and $ x_r $ are varying with time. Each user $ r \in R $, initially computes the price per unit flow by using the Eq. (\eqref{e2}) and it is willing to pay, $\mathcal{P}_r(t_i)$. 
Utility of each user $r$ is assumed as a strictly concave function of user's rate $ x_r(t_i) $ at that time ($ t_i $). At each time instant if user wants to access a link $e_{mn} \in E$ then, the cost will depend on the total data flow through the link at that time and is given by $\psi_{e_{mn}}(t_i)= \varsigma_{e_{mn}}(\sum_{r : e_{mn} \in E} x_r(t_i))$ where, $\varsigma_{e_{mn}}(\bullet)$ is growing if it comes in large number of user's path and $\varsigma_e(y)$ is given by 
\begin{equation*} 
\varsigma_{e_{mn}}(y)= c_{e_{mn}}.(y/C_{e_{mn}})^\omega 
\end{equation*}
where, $c_{e_{mn}}$ is constant and assumed with value one, $C_{e_{mn}}$ is the capacity of link $ e_{mn} \in E$. Now, consider the following system of differential equation for getting optimal data rate ($ x^* $)\\
\begin{equation}
\frac{dx_r(t_i)}{dt_i}= \vartheta_r(\mathcal{P}_r(t_i)-x_r(t_i)\sum_{e_{mn} \in r}\psi_{e_{mn}}(t_i)) \label{e4}
\end{equation}

Here, $ \vartheta_r $ is proportionality constant. Each user first computes it's willingness to pay as $\mathcal{P}_r(t_i)$ then, it adjusts its rate based on the response provided by the links in the network and trying to balance its willing to pay and total price. Eq. (\eqref{e4}) consists of two components: a steady increase in the rate proportional to $\mathcal{P}_r(t_i)$ and steady decrease in the rate proportional to the response $ \psi_{e_{mn}}(t_i) $ provided by the network.
\section{Time varying communication network model and routing strategies}
Network topology and routing are responsible for congestion and network efficiency. An optimal network structure is generated using an efficient TVCN model. Proposed model considers all the aspect of growth and alteration (rewiring \& removal) of links in the network. In the established network, users establish dedicated connection between the nodes with whom they want to communicate. These connections are able to manage traffic efficiently, which results minimized congestion in the network.
\subsection{Time varying communication network model}
Time varying data communication networks are designed by maintaining specific set of rules and nodes are connected via directed links. Directed links may have two categories: links which come out from a node (out-flowing) and other links are incidental to a node (in-flowing). Degree distribution of both out-flowing and in-flowing links follow a heavy tail power law distribution with distinct power law exponents. In the proposed model, the preferences are assigned to in-flowing links. The newly incoming node will be attached to the node endorsed by the highest number of nodes \cite{tadic,bara}. 
\subsubsection{Network Dynamics}
A communication network model is proposed  by using the Barabasi-Albert (BA) model \cite{bara} and the model using the dynamics of directed graphs \cite{tadic}. The proposed directed network model has following properties: 
\begin{enumerate}
\item Network consists of directed links for connecting nodes. 
\item Network expansion, link rewiring and removal are done at unique time scale.
\end{enumerate}
At each time instant $ t_i \in T $, a new node $ n_i $ is added to the network (expansion) and a number $ X (\leq n_0) $ is selected for network expansion, rewiring and removal, where $ n_0 $ is initial number of nodes present in the seed network. Links are divided into three categories: newly added, rewired and removed links. Distribution of the links are done using the given set of rules \cite{SK}.
\\
\textbf{Notations:}
\begin{enumerate}[(a.)]
\item $ \beta =$ Fraction of the evolving (appear from the new incoming node and appear/disappear in the existing network) links at any time instant. It informs about the establishment of new connections from the new incoming node $ n_t$ at time $ t $ ,  $ 0 < \beta < 1 $.
\item $ \gamma= $ Fraction of the links, those are rewired in the existing network, $0.5 < \gamma \leq  1 $.
\end{enumerate}
Using the above notation, following set of rules may be defined:
\begin{enumerate}[(i)]
\item Total number of new out-flowing links from the new appeared node, $ i (=t) $ with the nodes existing in the network at $ (t-1) $ based on preferential attachment is given by,
\begin{equation*}
 f_{add} (t) = \beta X 
\end{equation*}
\item Few links are rewired in the existing network. $ \gamma $ fraction of the available $ X $ links are chosen for rewiring,
\begin{equation*}
 f_{rewire}(t) = \gamma (X - f_{add}(t)) = \gamma (1-\beta) X 
\end{equation*}
\item Last part of the remaining segment of $ X $ are used for deleting most infrequently used links.
\begin{equation*}
f_{delete}(t) = X- f_{add}(t) - f_{rewire}(t) = (1-\gamma)(1-\beta) X
\end{equation*} 
\end{enumerate}
The parameter $ \delta $ is the ratio of altered and added links in the model and is given by,
\begin{equation}
\begin{aligned}
\delta ={} & \frac{f_{delete}(t)+ f_{rewire}(t)}{f_{add}(t)} \\
 	 	={} & \frac{(1-\gamma)(1-\beta) X + \gamma (1-\beta) X}{\beta X} \\
 	={} & \frac{(1-\beta)}{\beta} \label{e5}
 	\end{aligned}
\end{equation}
Here, $ \delta $ is independent of the number $ X $ and is known as \textbf{correlation parameter}.

While studying the behavior of communication networks, the concept of preferred linking to a node is driven by the demand of the node to outflow the data into the network. Only few influential nodes have the right to rewire their links. Nodes rewire their infrequently used links and connect it to preferred nodes for data communication. The preference is given to the infrequently used links for the removal of links. Hence, the links attached to a node having lower degree will be preferred for removal. 
\subsubsection{Degree Distribution}
The scale free behavior is found, after studying the scaling properties of the time varying network model. A mean field theoretical approach \cite{bara} is used to anticipate the growth dynamics of distinct nodes, which is later applicable for analytical computation of connectivity distribution and scaling exponents. Network is growing with time hence, some of the previously existing nodes will get more time to acquire links and are being the holders of highest degrees. Degree $ k_i $ of the node $ i $ is changing continuously with time, so probability $ \Pi(k_i)  $ is interpreted as rate of change of $ k_i $. At each time stamp, fractions $ \beta $ and $ \gamma $ decide the number of links chosen for expansion (addition), rearrangement and removal, and is deciding parameters of degree of the node $ i $. Analysis is given using following steps:
\begin{enumerate}
\item A fraction $ \beta $ of the number $ X $ links are newly added links at time $ t $.
\begin{equation}
\left( \frac{dk_i}{dt}\right)_{add} = \beta X \frac{k_i}{\sum_j k_j} \label{e6}
\end{equation}
Effect of newly added links on the degree of node $ i $ is written on the left hand side of the equation and on the right side $ \beta X $ links are formed using preferential attachment.
\item A fraction $ \gamma (1-\beta) X $ links are re-arranged at time $ t $. 
\begin{equation}
\left( \frac{dk_i}{dt}\right)_{rewire} = \gamma (1-\beta) X \left[ \frac{1}{n_0 + t} + \left( 1- \frac{1}{n_0 + t} \right) \frac{k_i}{\sum_j k_j} - \frac{1}{n_0 + t} \left( 1-\frac{k_i}{\sum_j k_j}\right)  \right] \label{e7}
\end{equation}
When few links are rewired then, change in degree of node $ i $ depends on three terms: first term shows random selection of nodes, second term corresponds to linking with other existing nodes having high preferential attachment probability and third term shows removal of link having low preferential attachment value. 
\item A fraction $ (1-\gamma) (1-\beta) X $ links are removed from the network at time $ t $.
\begin{equation}
\left( \frac{dk_i}{dt}\right)_{delete} = -(1-\gamma)(1-\beta)X\left[\frac{1}{n_0+t} + \left(1-\frac{1}{n_0 + t} \right) \left( 1-\frac{k_i}{\sum_j k_j}\right)\frac{1}{n_0+t}\right] \label{e8}
\end{equation}
Removal of links affect the degree of node $ i $ and it is shown in above equation. First term shows random selection of a node and that link will be removed which has low preferential attachment value.
\end{enumerate}

At time $ t $, the sum of degrees of nodes in the network will be:
\begin{equation*}
\sum_j k_j = 2t \left[\beta X + (1-\beta)\left(\gamma X - (1-\gamma) X \right)\right]
		= 2Xt\left[ \beta + (1-\beta)(2\gamma-1)\right]  
\end{equation*} 
Let $ c = \beta + (1-\beta)(2\gamma-1) $.\\
Now, combining the eqns. \eqref{e6} to \eqref{e8} we get the change in degree of node $ i $ with respect to time $ t $.

\begin{equation}
\begin{aligned}
\frac{dk_i}{dt} ={} & \frac{\beta k_i}{2ct} + \frac{\gamma(1-\beta)k_i}{2ct} \nonumber \\
& - (1-\beta)(1-\gamma)X \left[ \frac{2}{n_0+t} - \frac{1}{(n_0+t)^2} - \frac{k_i}{2ct(n_0+t)} + \frac{k_i}{2ct(n_0+t)^2} \right]\\
={} & \left( \frac{\beta}{2c} + \frac{\gamma(1-\beta)}{2c} \right)\frac{k_i}{t} - \left( (1-\beta)(1-\gamma)X \right) \frac{1}{t}\\
&\mbox { (for large  t ) }
\end{aligned}
\end{equation}
Let us assume $ \theta_1 =  \frac{\beta}{2c} + \frac{\gamma(1-\beta)}{2c}  $ and 
$ \theta_2 = -(1-\beta)(1-\gamma)X $\\
Hence, the above equation can be re-written as\\
\begin{equation}
\frac{dk_i}{dt} = \theta_1 \frac{k_i}{t} + \theta_2 \frac{1}{t}\label{e9}
\end{equation}
The solution of the Eq. (\eqref{e9}) is derived by taking initial condition that node $ i $ appears at time $ t_i $ with $ X $ connections $ k_i(t_i) = X $ and is given as:
\begin{equation*}
\begin{aligned}
k_i(t)={} & -\frac{\theta_2}{\theta_1} + \left( k_i(t_i) + \frac{\theta_2}{\theta_1} \right) \left( \frac{t}{t_i} \right) ^ {\theta_1} \\
   ={}   & -\frac{\theta_2}{\theta_1} + \left( X + \frac{\theta_2}{\theta_1} \right) \left(\frac{t}{t_i} \right) ^ {\theta_1}
\end{aligned}
\end{equation*}
At each time stamp, a node is added into the network hence, Probability density of $ t_i $ is:
\begin{equation*}
P_i(t_i) = \frac{1}{n_0 + t}
\end{equation*}
The probability that a node has total $ k_i(t) $ connections and which is smaller than $ k $, $ P(k_i(t) < k) $, can be written as:
\begin{equation*}
\begin{aligned}
P(k_i(t) < k) = {} & P(-\frac{\theta_2}{\theta_1} + \left( X + \frac{\theta_2}{\theta_1} \right) \left(\frac{t}{t_i} \right) ^ {\theta_1} < k)\\
={}  & P \left( t_i > \left( \frac{X + \frac{\theta_2}{\theta_1}}{k+\frac{\theta_2}{\theta_1}} \right) ^ {\frac{1}{\theta_1}} t \right)\\
={} & 1 - P \left( t_i \leq \left( \frac{X + \frac{\theta_2}{\theta_1}}{k+\frac{\theta_2}{\theta_1}} \right) ^ {\frac{1}{\theta_1}} t \right)
\end{aligned}
\end{equation*}
The probability density of $ k $ is $ P(k) $ and can be written as:\\
\begin{equation}
\begin{aligned}
P(k) = {} & \frac{\partial P(k_i(t) < k) }{ \partial k}\\
={} & \frac{t}{\theta_1(n_0+t)} \left( X + \frac{\theta_2}{\theta_1} \right) ^ {\frac{1}{\theta_1}}\left( k + \frac{\theta_2}{\theta_1} \right) ^ {-(1+\frac{1}{\theta_1})} \label{e10}
\end{aligned}
\end{equation}
Now, assume $ \alpha = -(1+\frac{1}{\theta_1}) $. \\
The value of the scaling exponent of the real world network lies between $ 2 $ and $ 3 $, hence, for dynamic communication network, exponent ($ \alpha $) must lie within $ 2 < \alpha \leq 3  $. Constraint on $ \alpha $ will be fulfilled if $ 0.5 < \theta_1 < 1 $. As $ P(k) $ is always positive so, $ X + \frac{\theta_2}{\theta_1} $ should be positive. 
\subsection{Routing Strategies}
Each node in the communication network generates packets with a certain rate and can deliver packets according to the capacity of that particular node. Capacity of the network ($C$) is calculated by summing up the capacity of all the individual nodes in the communication network. Aggregate values of the packet generation rate of all the nodes are termed as the load of the network.
\begin{equation*}
\begin{aligned}
L={}& \sum_{v = 0}^{N-1} l_v \\
& \mbox{ here, } l_v = \mbox{ load at  node } v 
\end{aligned}
\end{equation*}
When, load ($L$) exceeds the capacity ($C$) of the network then, it will become congested with the packets. If all the packets are sent through the shortest paths to their destinations then, some nodes  may appear frequently in the formation of shortest path and it will become congested. In the large communication networks, multiple users want to establish a connection between a source ($S$) and destination ($D$). There may exist multiple shortest paths $ \sigma_z(S \rightarrow D) $ from $ S $ to $ D $, where $ z = {1,2, ..., \chi} $ and $ \chi $ is total number of shortest paths between $ S $ and $ D $. In the shortest path, a node may lie with high betweenness centrality. Sending data through the shortest path will lead to the congestion in the network (most of the user pairs will try to follow the same path). Therefore, it is important to investigate a shortest path $ \sigma(S \rightarrow D) $ between the user pairs such that overall betweenness centrality of the nodes appears in the path
\begin{equation*}
\begin{aligned}
\theta_g[\sigma_z(S \rightarrow D)] ={}& \sum_{v: v \in \sigma_z(S \rightarrow D)} g_v \\
& \mbox{ here, } g_v = \mbox{ betweenness of node  } v
\end{aligned}
\end{equation*} 
should be minimum. Hence, we want to find out a path whose nodes should not be the part of large number of  other shortest paths (low betweenness centrality). Hence, it is defined by,
\begin{equation*}
min \mbox{  } \{\forall z : \theta_{g}(\sigma_z(S \rightarrow D))\}
\end{equation*}
Apart from routing strategy using betweenness centrality of the nodes, a local strategy is used by considering influence of the neighbor nodes. If a heavily loaded node is connected with some other nodes then, it will make all the succeeding nodes congested. Heavily congested preceding nodes have negative impact on successor nodes. The impact of one node to others is measured by Eigen vector centrality of the node $ v $,
\begin{equation*}
x_v = \frac{1}{\kappa} \sum_{j \in N} a_{vj} x_j
\end{equation*} 
Here, $ \kappa $ is largest Eigen value of the matrix $ A $ and if node $ v $ and $ j $ are connected then, $ a_{vj} = 1 $. If a node is most Eigen vector central, then it will be highly congested and hence, least reputed. So, among all the shortest path, that path will be reputed if aggregate value of the reputation of the path $ \theta_r[\sigma(S \rightarrow D)] $
\begin{equation*}
\begin{aligned}
\theta_r[\sigma_z(S \rightarrow D)] ={}& \sum_{v: v \in \sigma_z(S \rightarrow D)} \frac{1}{x_v}\\
& x_v = \mbox{ Eigen vector centrality of node  } v
\end{aligned}  
\end{equation*}
is maximized and and the function is defined by,
\begin{equation*}
max \mbox{  } \{\forall z : \theta_{r}(\sigma_z(S \rightarrow D))\}.
\end{equation*}
Optimal path between $ S $ and $ D $ 
\begin{equation*}
\begin{aligned}
\theta_{gr}\sigma_z(S \rightarrow D)] ={}& \sum_{v: v \in \sigma_z(S \rightarrow D)} {g_v}{x_v}
\end{aligned}
\end{equation*}
is calculated by considering both betweenness centrality and reputation. A path is optimal if it's elements will have least betweenness central as well as reputed and is given by,
\begin{equation*}
min \mbox{  } \{\forall z : \theta_{gr}(\sigma_z(S \rightarrow D))\}.
\end{equation*}

\begin{figure*}[!htb]
\begin{center}
\includegraphics[width=\linewidth, height=3 in]{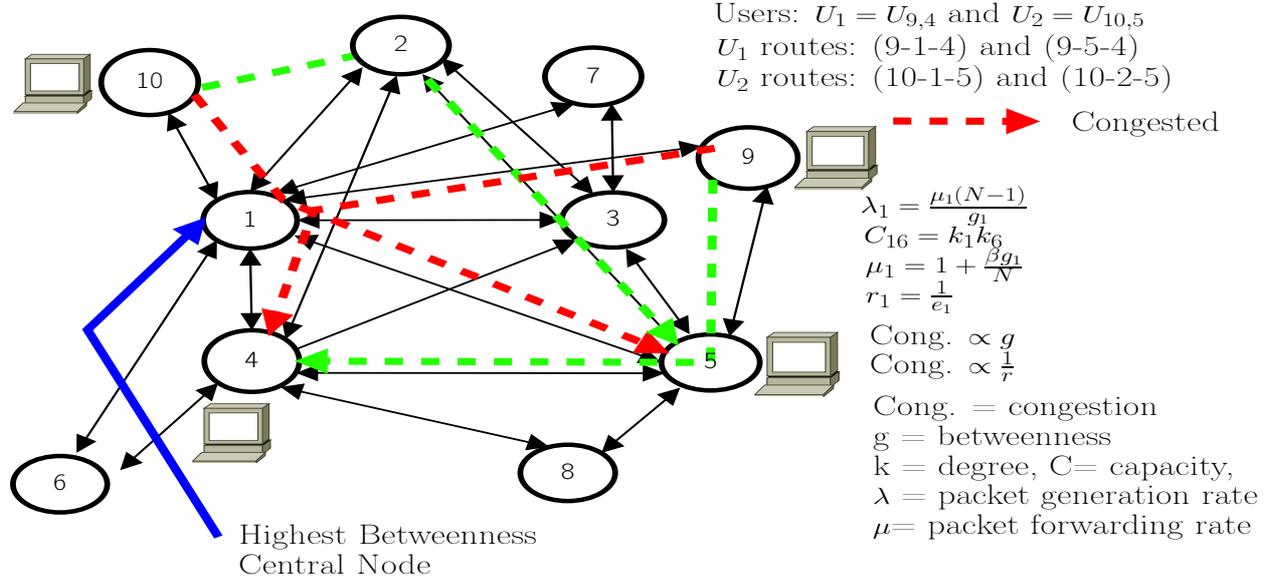}
\caption{Block diagram}
\label{f1}
\end{center} 
\end{figure*}
A snapshot of the proposed method is discussed in Fig. \ref{f1}. Two users, $ U_1 $ and $ U_2 $ want to send data from node $ 9 $ to node $ 4 $ and from node $ 10 $ to node $ 5 $, respectively. Their respective routes are $ 9-1-4, 9-5-4 $ and $ 10-1-5, 10-2-5 $. Both the users avoid congested routes hence, $ U_1 $ avoids the route $ 9-1-4 $ and $ U_2 $ prefers $ 10-2-5 $. From Fig. \ref{f1}, it is evident that node $ 1 $ is highly congested node in the network, so both users avoid to send data through node $ 1 $ and select other possible options available for sending data to their destined node.  Packet generation rate (PGR) of a node depends on the packet forwarding rate and betweenness centrality of the node. Packet forwarding rate depends on the capacity of a node, that affects bandwidth of a link.

Capacity of a link $ C_{e_{mn}} $ is dependent on the degrees $ k_m $ and $ k_n $ of node $ m $ and $ n $. It can be approximated by a power-law dependence \cite{onnela}
\begin{equation*}
C_{e_{mn}} = b (k_m k_n)^\alpha
\end{equation*}
$ \alpha $ is the degree influenced exponent which depends on the type of networks and $ b $ is a positive quantity. Three possible relations between capacity $ C_{e_{mn}}  $ and exponent $ \alpha $: (i) if $ \alpha > 0 $ then, data is transmitted through high degree nodes (ii) when $ \alpha < 0 $ then, data will be sent through low degree nodes and (iii) $ \alpha =0 $ shows degree independent transmission. The value of $ \alpha $ in the proposed model is taken positive value with constant $ \alpha = 1 $ and $ b $ is also equal to $ 1 $. Reputation and betweenness centrality are deciding parameter for evaluating congestion. Reputation of a node is inverse of the Eigen vector centrality of a node. While, congestion is proportionally related to betweenness centrality and inversely related with reputation.    

Algorithmic steps are given for expansion, rewire and removal of links in the TVCN (Algorithm \ref{al1}).\\

\begin{algorithm}[htb!]
\begin{algorithmic}[1]
\STATE \textbf{Input:}  A number of nodes $ (n_0) $ for creating seed network, $ n (\leq n_0)$, $ \beta $, $ \gamma $ and $ T $.
\STATE \textbf{Output:} Time varying data communication networks.  
\WHILE{$ t  \leq T $}
\STATE Expand the network with one node at each time instant $ t $.
\FOR{$ x $: 1 to $ f_{add}(t) $}
\STATE  Attach the newly added node with node having highest inflowing link attachment probability in the existing network.
\ENDFOR
\FOR{ $ y $ : 1 to $f_{rewire}(t)$}
\STATE Remove the infrequently used link of an influenced node and attach to the node having higher inflowing link probability.
\ENDFOR
\FOR{$ z $ : 1 to $f_{delete}(t)$}
\STATE Remove the infrequently used link of a randomly selected node.
\ENDFOR
\ENDWHILE\label{endtimer}
\end{algorithmic}
\caption{Network Evolution}
\label{al1}
\end{algorithm}

Steps involved in Algorithm \ref{al1}, are used to design TVCN. In the time varying network, various users want to communicate data from a source $ S $ to destination $ D $. User gives information about the source and destination, and accordingly $ S \rightarrow D $ pairs are generated. Increments in the number of users lead to the network into congested state. Therefore, our aim is to find efficient routing paths such that maximum number of users are getting benefited with a unique stable value of the optimal data sending rate $ x_r^* $ and corresponding convergence vectors will be $ x^*={x_r^*, r \in R}$ using equation Eq. (\ref{e4}). Shortest path $ \sigma(S \rightarrow D) $ with lowest betweenness centrality and highest reputation $( \theta_{gr}[\sigma(S \rightarrow D)]) $ are assumed as efficient. After finding such paths of all the users, optimal data sending rate of each user is calculated by using rate allocation equation Eq. (\ref{e4}). Detailed description of selecting shortest path with lowest (highest) betweenness centrality and highest (lowest) reputation along with optimal rate is given in Algorithm \ref{al2}. 

\begin{algorithm}[H]
\begin{algorithmic}[1]
\STATE \textbf{Input:} All source destination $ S-D $ pairs $ \mathcal{N}_{SD} $ in the network, designed through Algorithm \ref{al1}, $ a $ and $ b $ such that $ a > 0 $ \& $ 0 < b < 1 $.
\FOR {$ i:=  1 \mbox{ to } length(\mathcal{N}_{SD})$}
\STATE Evaluate all shortest paths $ \chi_i $ of user $ i $.
\IF{ $  \chi_i  > 1 $}
\FOR {$ m:= 1 \mbox{ to }  z $}
	\STATE Calculate $ \theta_{gr}[\sigma_m(S \rightarrow D)] $.		
\ENDFOR
\STATE Select shortest paths $ \chi_i(S \rightarrow D) $  having maximum and minimum value of $ \theta_{gr}[\sigma_m(\chi_i(S \rightarrow D ))]$.
\ELSE
\STATE Calculate $ \theta_{gr}[\sigma(S \rightarrow D)] $.   
\ENDIF
\FOR {$ j:= 1 \mbox{ to } N  $} 
\STATE Frequency of occurring of node $ j $, $ \omega_j $ in user's path decides initial data rate of each node.
\STATE $ x_j = \frac{C_j}{\omega_j} $;
\ENDFOR
\ENDFOR
\FOR { $ i := 1 \mbox{ to } length(\mathcal{N}_{SD}) $ }
	 \FOR { $ d := 1 \mbox{ to } length(\mathcal{N}_{SD}(i))$}
        \STATE Update network feedback $ \psi_d $ for each element $ d $.
     \ENDFOR
        \STATE $ x_r = min(x_{\mathcal{N}_{SD}(i)}) $;
        \STATE $ A(r) = rand(1,10) $;
        \STATE $ \mathcal{P}_r = x_r * (\frac{a}{x_r+ b}) $;
        \STATE $ \psi_r = \psi_d \{ \forall d : d \in \mathcal{N}_{SD}(i)\}$;
\ENDFOR
\STATE Use the value of $ x_r $, $ A(r) $, $ \mathcal{P}_r $ and $ \psi_r $ to find the rate of convergence of each user.  
\end{algorithmic}
\caption{Finding shortest path having lowest (highest) betweenness centrality with highest (lowest) reputation and optimal rate for each user }
\label{al2}
\end{algorithm}
\section{Simulation and results}
For dynamics of TVCN model, the simulation starts by establishing the infrastructure of the network followed by algorithmic steps in Algorithm \ref{al1} and the rules are given in the Section 3.1.1. The proposed TVCN model is one of the example of real world networks, hence, degree distributions must follow a heavy-tailed power law distribution: $ P(k) \sim k^{-\alpha} $, where $ 2 < \alpha \leq 3 $, known as scale-free networks. Scale-free networks are non-homogeneous, numerous nodes  with few links coexist  
with a few hub nodes, connected with thousands or  even millions of links. In this paper, the parameters are set to be seed node $ n_0 = 5 $, number $ X = 5 $, fraction of newly added links $ \beta $ range in $ (0,1) $, fraction of rewired links $ \gamma $ is in the range of $ (0.5, 1) $, with network size ranging from $ N = 10^2 $ to $ N = 10^4 $. Any nodes can be included in the user's $ S-D $ sets or may participate in routing also. Data forwarding capacity of a node $ n $, $ C_n $ is equal to the in-flowing degree $ k_n $ of the node $ n $. Capacity of a link $ C_{e_{mn}} $ is obtained by multiplying the degree $ k_m $ and $ k_n $ of end nodes $ m $ and $ n $ respectively. At each time stamp, degree of the nodes will be different, hence capacity of the nodes as well as links change accordingly.  

In the Section 3.1.1, the value of correlation parameter, $ \delta = \frac{1-\beta}{\beta} $, where, $ 0 < \beta < 1 $ , implies that $ \infty < \delta \leq 0 $. The value of $ \delta $ will be minimum with value $ 0 $ or maximum with value $ \infty $, for maximum value of $ \beta = 1 $, or minimum value of $ \beta = 0 $. The following cases are possible for the value of the $ \delta $, depending on altered (rewired \& removed) and newly added links in the network, 
\begin{enumerate}[(a)]
\item $ \delta = 1 $, when both the values of altered (rewired \& removed) and added links are same.
\item $ \delta < 1 (\beta > 0.5) $, more number of  new links are getting added than altered.
\item $ \delta > 1 (\beta < 0.5) $, more number of links are altered than added. 
\item $ \delta = 0 (\beta = 1) $, only expansion, no alteration in the existing network.
\end{enumerate}
\begin{figure*}[!htb]
\begin{center}
$\begin{array}{cc}
\includegraphics[width=0.45\linewidth, height=2.5 in]{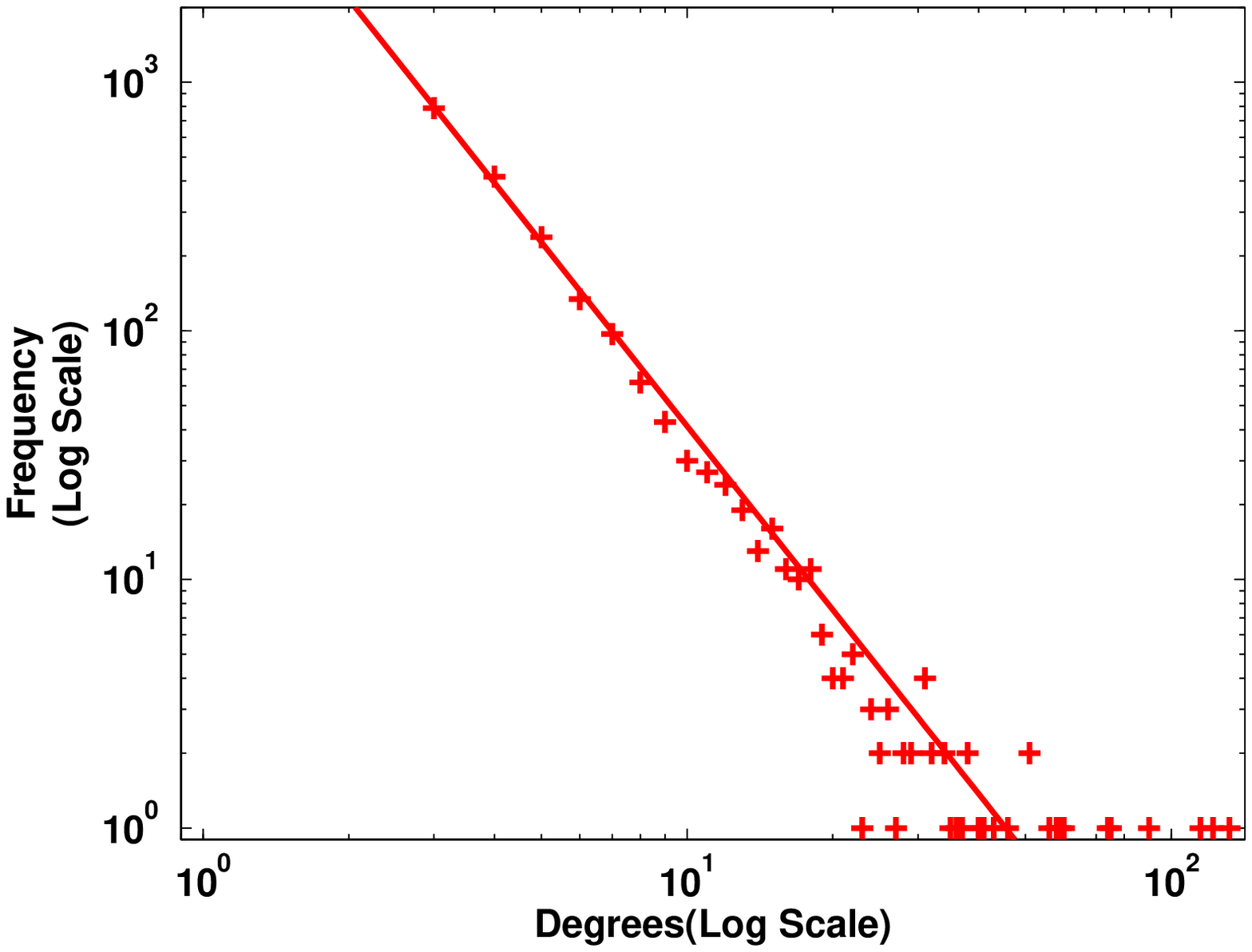} &
\includegraphics[width=0.45\linewidth, height=2.5 in]{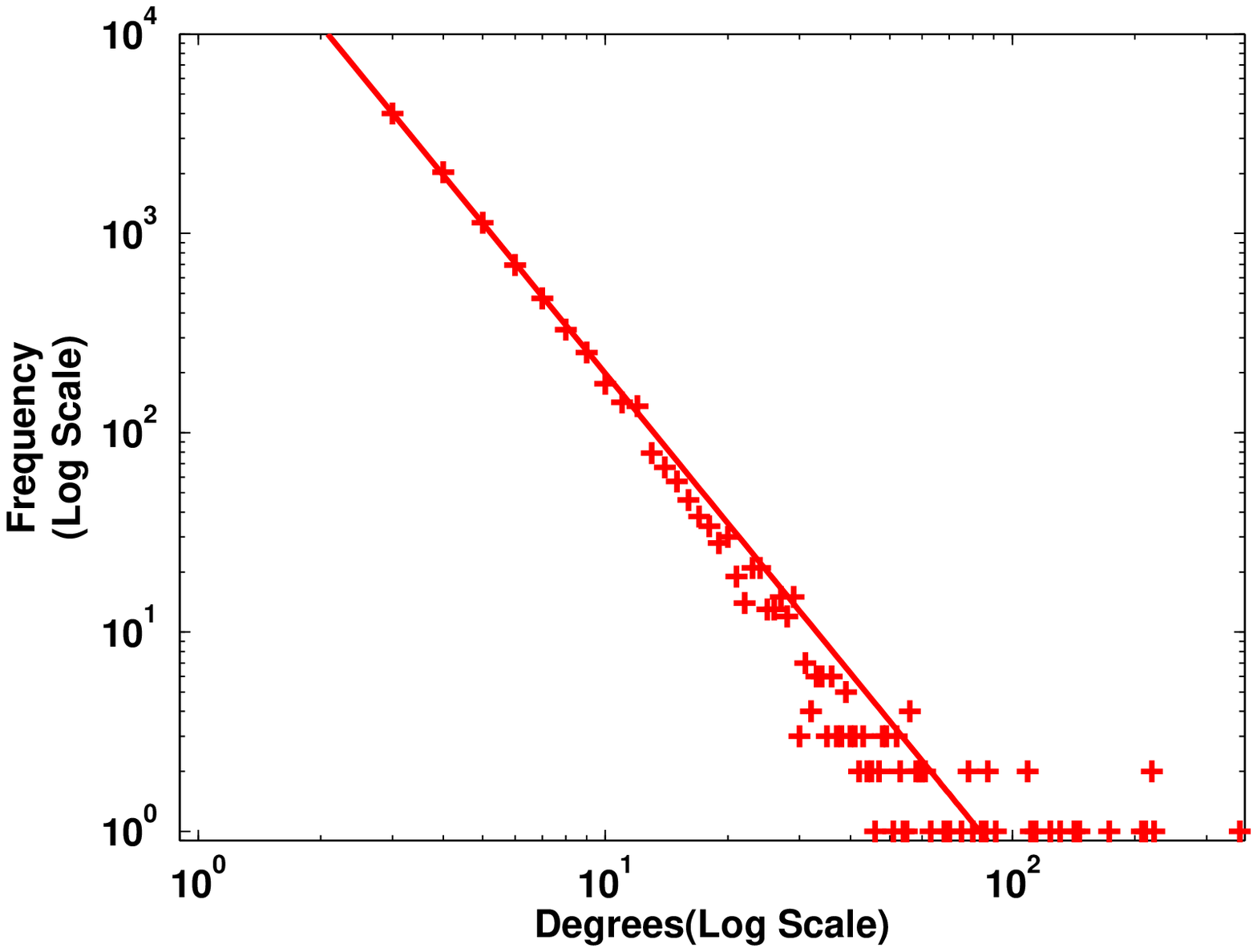}
 \\
\mbox{(a)} & \mbox{(b)}  \\
\includegraphics[width=0.45\linewidth, height=2.5 in]{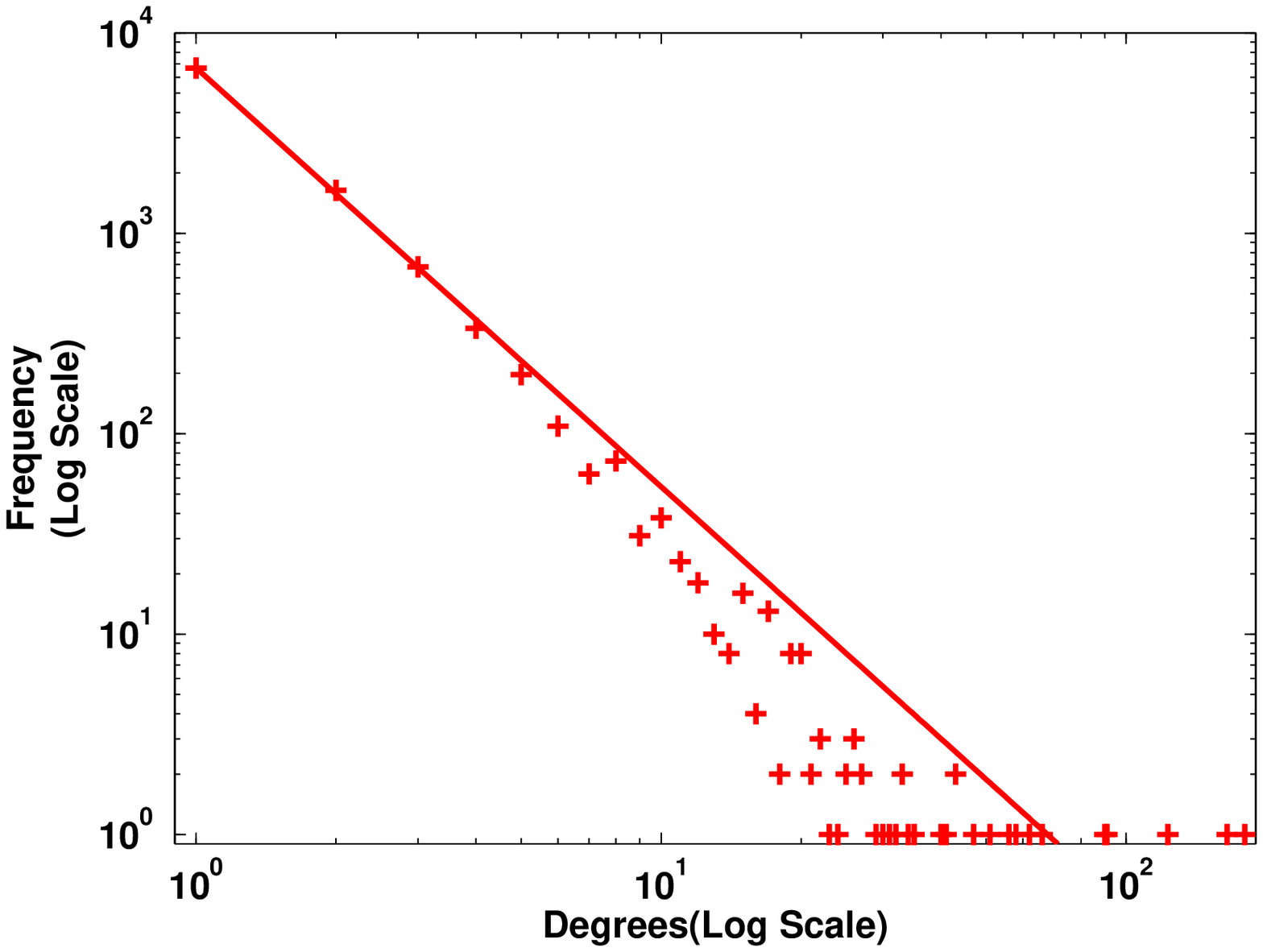} &
\includegraphics[width=0.45\linewidth, height=2.5 in]{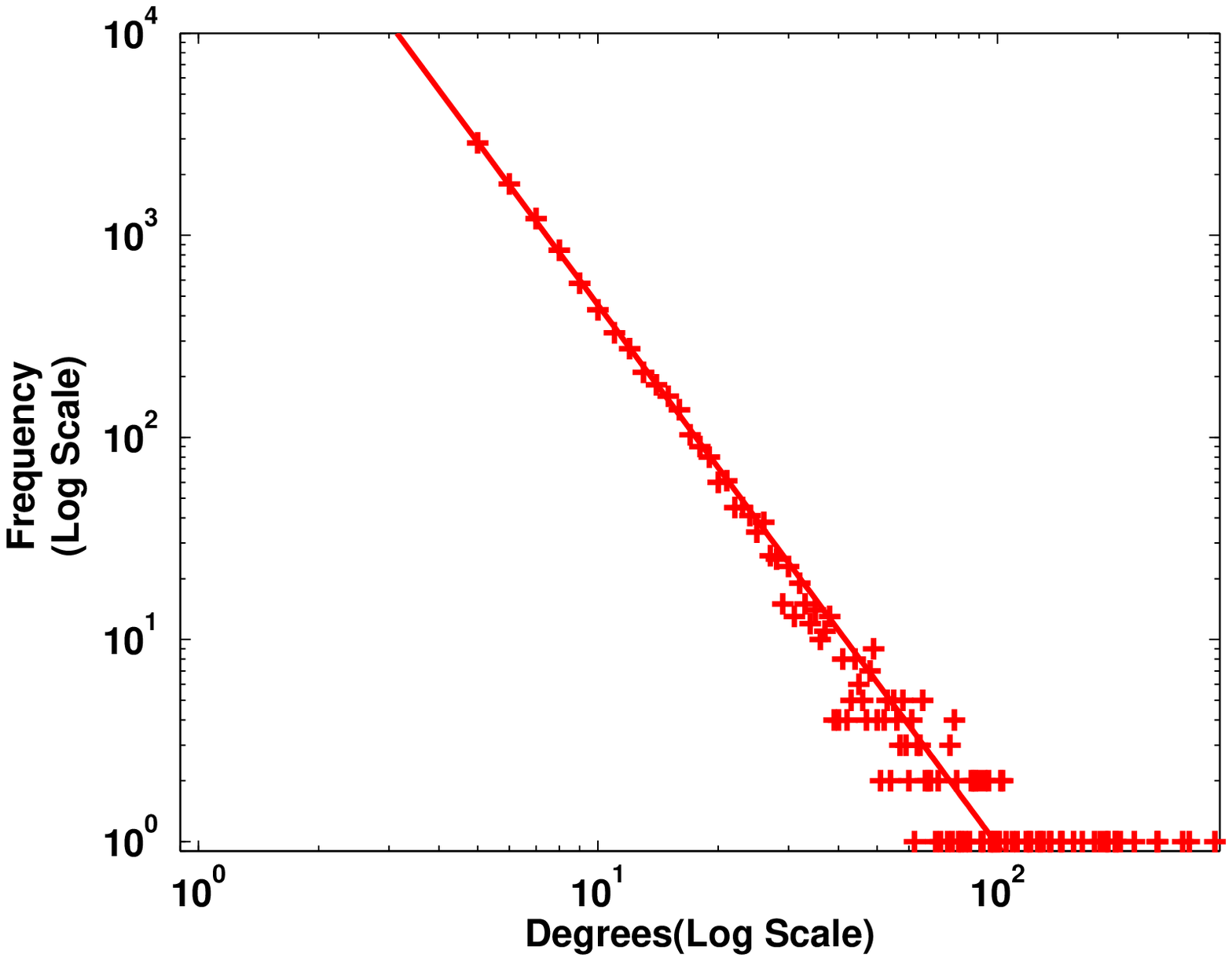} \\
\mbox{(c)} & \mbox{(d)}
\end{array}$
\caption{Degree distribution of the network when number of nodes are average ratio of altered and newly added links are \textbf{(a)} $ N = 10^4 \mbox{ , } \delta = 1.00   $, \textbf{(b)} $ N = 10^4 \mbox{ , } \delta = 0.67  $,  \textbf{(c)} $ N = 10^4 \mbox{ , } \delta = 3.00 $ and \textbf{(d)} $  N = 10^4 \mbox{ , } \delta = 0.00 $}
\label{f2}
\end{center} 
\end{figure*}

The degree distribution is plotted of the generated TVCN (Fig. \ref{f2} $ (a-d) $) for number of nodes, $ N = 10^4 $. Different values are considered for parameters $ \beta $, $ \gamma $ and $ \delta $ (Table \ref{tab1} $ (a-d) $). The degree distribution is found power law in nature with all the considered values of the parameters (Fig. \ref{f2}). A fraction $ (1-\beta) $ of the number $ X $ is used for alteration of links and $ \gamma $ part of the altered links are used for rewiring. To maintain the growing nature of TVCN, the value of $ \gamma $ is considered in the range $ (0.5, 1) $. Once the value of $ \beta $ is increased it implies that more number of new incoming links appear with the newly added node. Therefore, in the network a node may acquire more number of links. Hence, the tail of the distribution is found longer and increasing slope of degree distribution of TVCN (Fig. \ref{f2} $ (a -d) $). The power law exponent $ \alpha $ is also increasing ($ 2 < \alpha \leq 3 $) for all the possible values of correlation parameter $ \delta $ (Table \ref{tab1}), with the increasing value of $ \beta $. 

\begin{table}[!htb]
\caption{Values of power law exponent $ \alpha $, for different values of $ \delta $ at $ N = 10^4 $.} 
 \center
\begin{tabular}[1]{l| c| c| c}
\hline\hline
& \textbf{$ \beta $} & \textbf{$ \gamma $} & $ \alpha $
\\ [0.5ex]
\hline\hline
$ \delta = 1 $ & $ 0.50 $ & $ 0.50 $ & $ 2.486 $  \\[0.25ex]
\hline
$ \delta < 1 $ & $ 0.60 $ & $ 0.50 $ & $ 2.455 $ \\[0.25ex]
\hline
$ \delta > 1 $ & $ 0.25 $ & $ 0.70 $ & $ 2.065 $  \\[0.25ex]
\hline
$ \delta = 0 $ & $ 0.90 $ & $ 0.90 $ & $ 2.664 $ \\[0.25ex]
\hline
\end{tabular}
\label{tab1}
\end{table}


As per proposed mathematical model for calculating the degree distribution of the generated TVCN, power law exponent $ \alpha $ is dependent on the value of $ \theta_1 $ and may be written as $ \frac{1}{\alpha +1} $. For real world networks, range of $ \alpha $ is $ (2, 3] $ hence, $ \theta_1 $  must lie between $ 0.5 $ and $ 1 $. The density function is always positive, hence, $ X + \frac{\theta_2}{\theta_1} $ must be positive. For different values of $ \delta $, both conditions are fulfilled (Table \ref{tab2}).   
\begin{table*}[!htb]
\caption{User's optimal rate through the shortest routes having different betweenness centrality values (maximum and minimum), when number of nodes $ N=10^2 $} 
 \center
\begin{tabular}[1]{l| c| c| c| c| c} 
\hline\hline 
& $ \beta $ & $ \gamma $ & $ \theta_1 $ & $ \theta_2 $ & $ \left( X + \frac{\theta_2}{\theta_1} \right) $
\\ [0.5ex]
\hline\hline 
$ \delta = 1 $ & $ 0.5 $ & $ 0.6 $ & $ 0.75 $ & $ -1.25 $ & $ 3.336 $  \\
\hline
$ \delta < 1 $ & $ 0.6 $ & $ 0.5 $ & $ 0.667 $ & $ -1 $ & $ 4.351 $  \\
\hline
$ \delta > 1 $ & $ 0.25 $ & $ 0.7 $ & $  0.7045 $ & $ -1.125 $ & $ 3.4019 $\\
\hline
$ \delta = 0 $ & $ 0.9 $ & $ 0.5 $ & $ 0.5278 $ & $ -0.25 $ & $ 4.526 $ \\
\hline
\end{tabular}
\label{tab2}
\end{table*}

\begin{figure}[!htb]
\begin{center}
$\begin{array}{cc}
\includegraphics[width=0.50\linewidth, height=2.5 in]{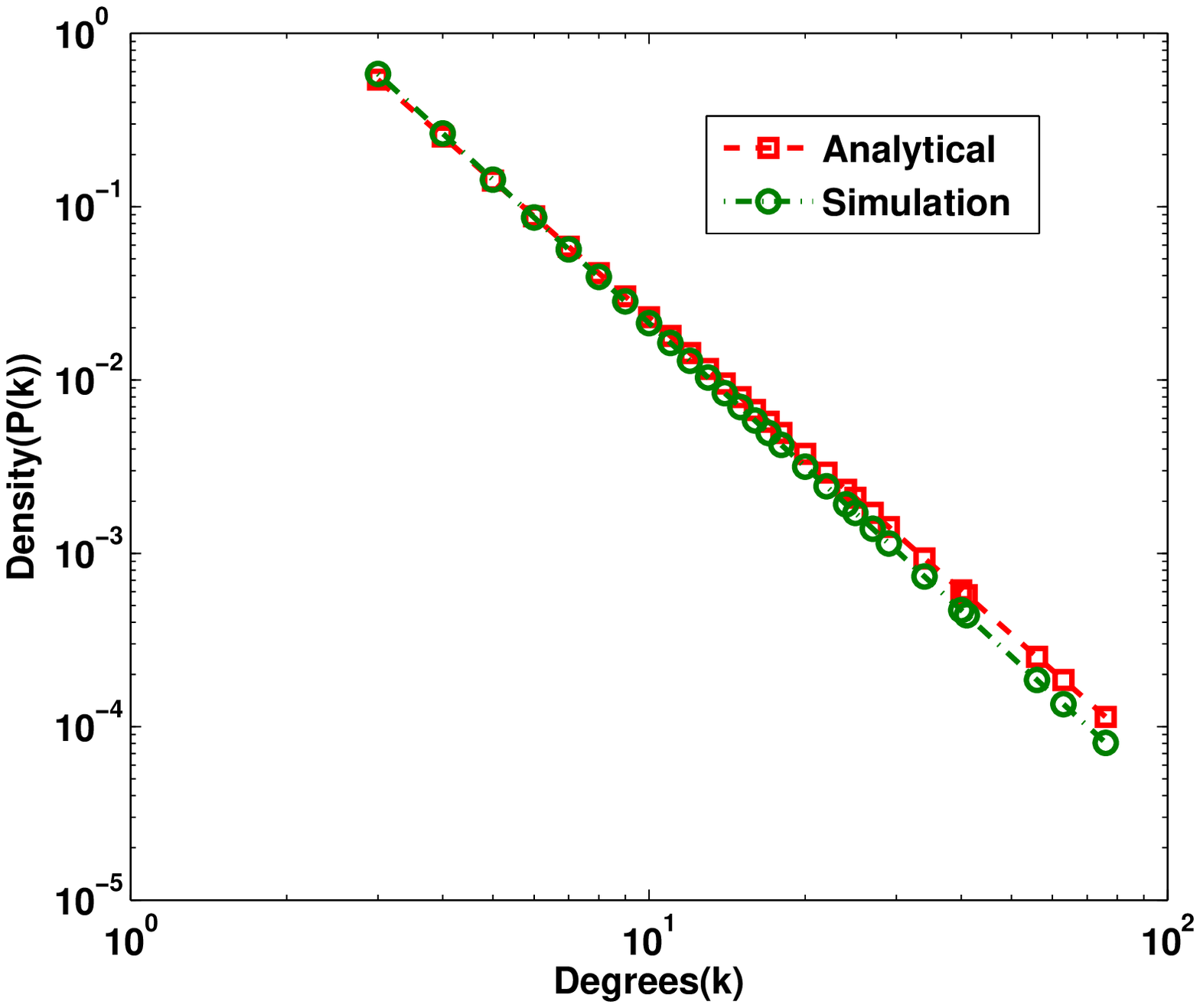} &
\includegraphics[width=0.50\linewidth, height=2.5 in]{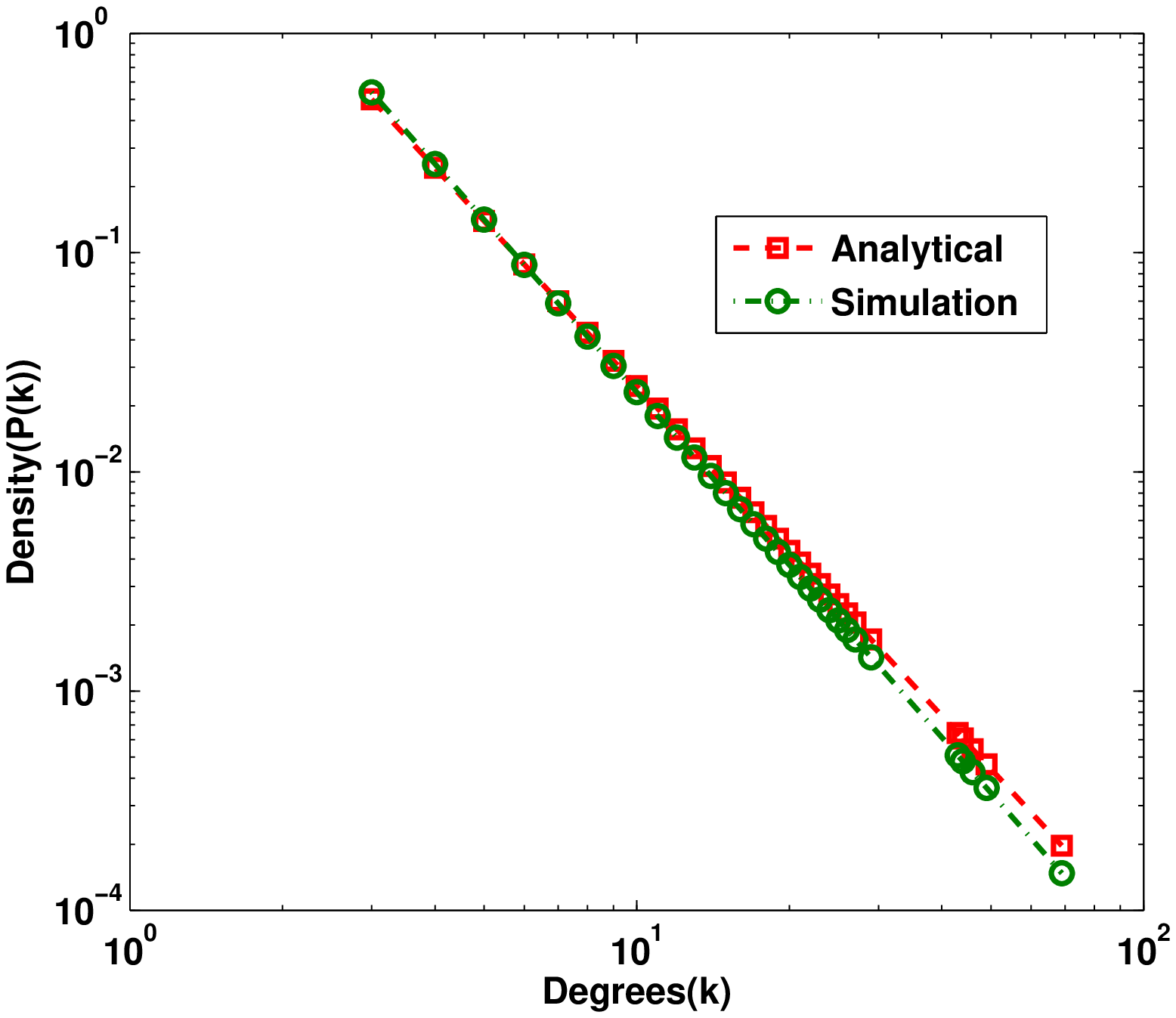} \\
\mbox{(a)} & \mbox{(b)}\\
\includegraphics[width=0.50\linewidth, height=2.5 in]{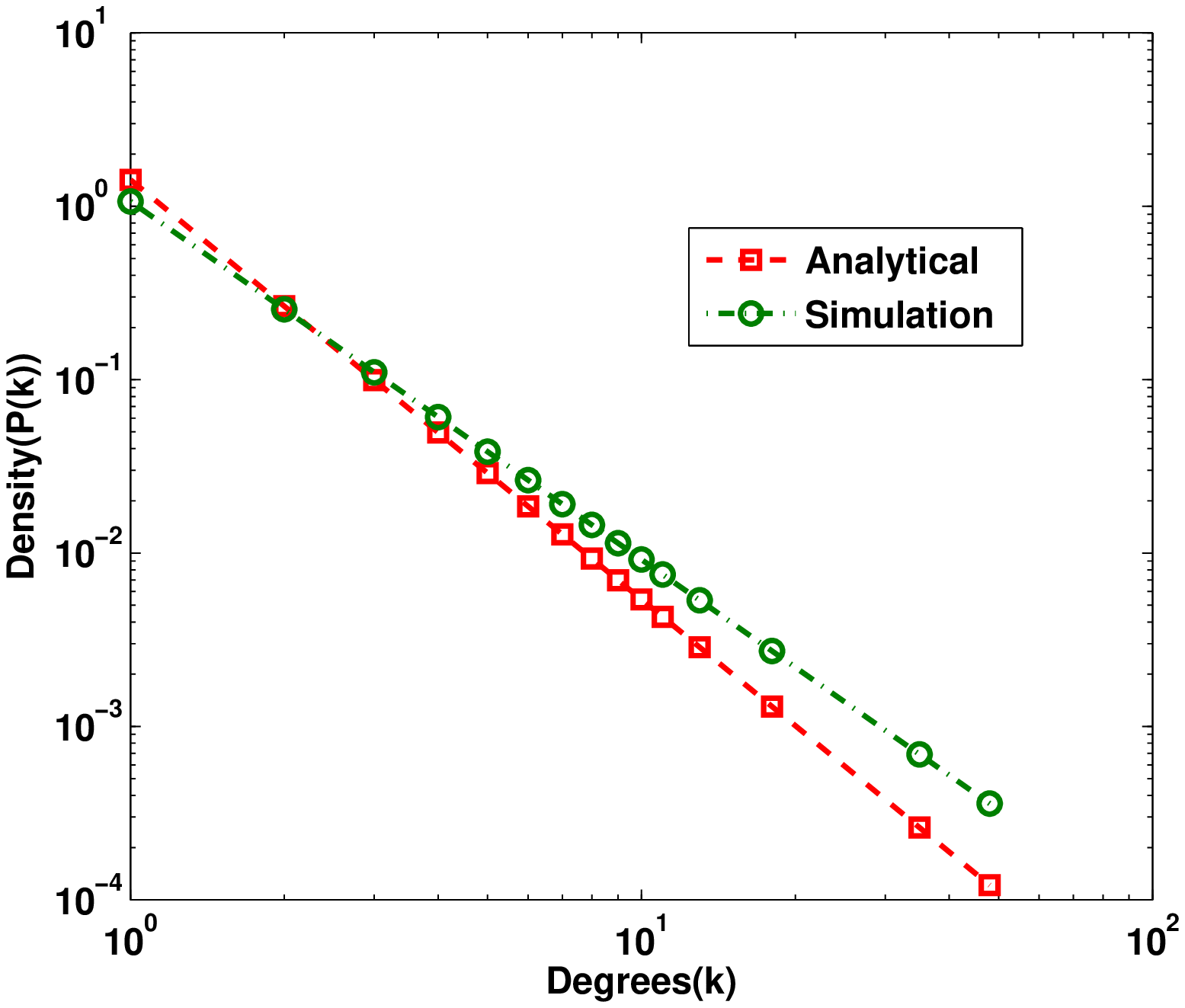} &
\includegraphics[width=0.50\linewidth, height=2.5 in]{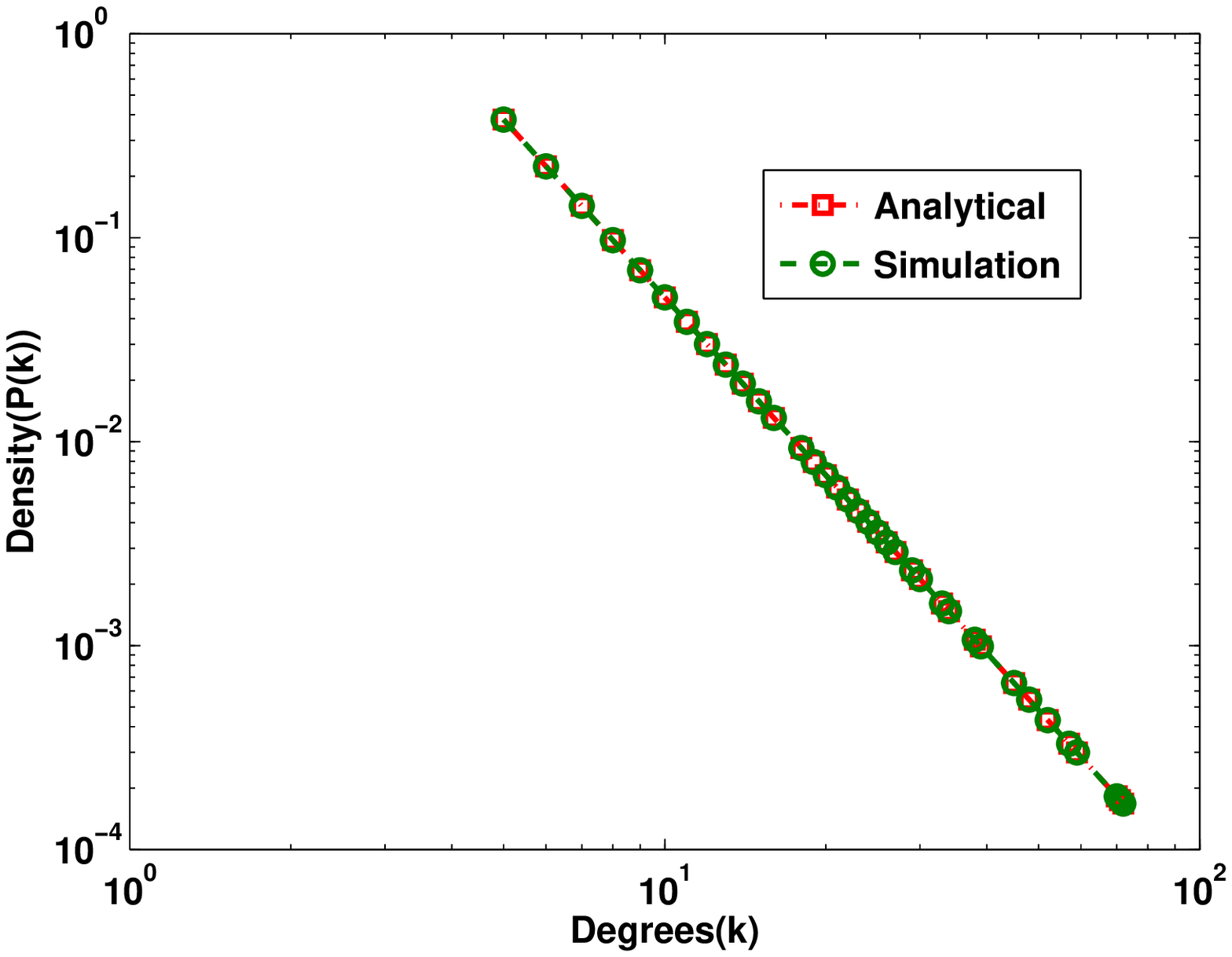} \\
\mbox{(c)} & \mbox{(d)}\\
\end{array}$
\caption{Simulated and analytical results of degree distribution of in-flowing links for the network of $ N = 5 * 10^2 $ nodes for $ \delta = 1 $ (a), $ \delta < 1 $ (b), $ \delta > 1 $ (c) and $ \delta = 0 $ (d).}
\label{f3}
\end{center} 
\end{figure}

The degree distribution of the in-flowing links in TVCN is calculated and compared with all the possible cases of the parameter $ \delta $, by using mean field theoretical approach and through simulation, for the network of size $ N = 5*10^2 $. The obtained values of the power law exponent $ \alpha $ are very proximate to each other (Fig. \ref{f3}) with corresponding values (Table \ref{tab4}). Hence, proposed mathematical model for degree distribution for TVCN is validated. 

\begin{table}
\caption{Comparative study of the values of power law exponent $ \alpha $ obtained through simulation and theoretical approach, for different values of $ \delta $ for network size $ N = 5 * 10^2 $.} 
 \center
\begin{tabular}[1]{l| c| c| c| c}
\hline\hline
& \textbf{$ \beta $} & \textbf{$ \gamma $} & $ \alpha $ & $ \alpha $
\\ [0.5ex]
& & & (Simulation) & (Theoretical)\\
\hline\hline
$ \delta = 1 $ & $ 0.50 $ & $ 0.50 $ & $ 2.750 $ & $ 2.619 $  \\[0.25ex]
\hline
$ \delta < 1 $ & $ 0.60 $ & $ 0.50 $ & $ 2.616 $ & $ 2.500 $ \\[0.25ex]
\hline
$ \delta > 1 $ & $ 0.25 $ & $ 0.70 $ & $ 2.021 $ & $ 2.419 $  \\[0.25ex]
\hline
$ \delta = 0 $ & $ 0.90 $ & $ 0.90 $ & $ 2.741 $ & $ 2.895 $ \\[0.25ex]
\hline
\end{tabular}
\label{tab4}
\end{table}

\begin{figure*}[!htb]
\begin{center}
\includegraphics[width=\linewidth, height=2.8 in]{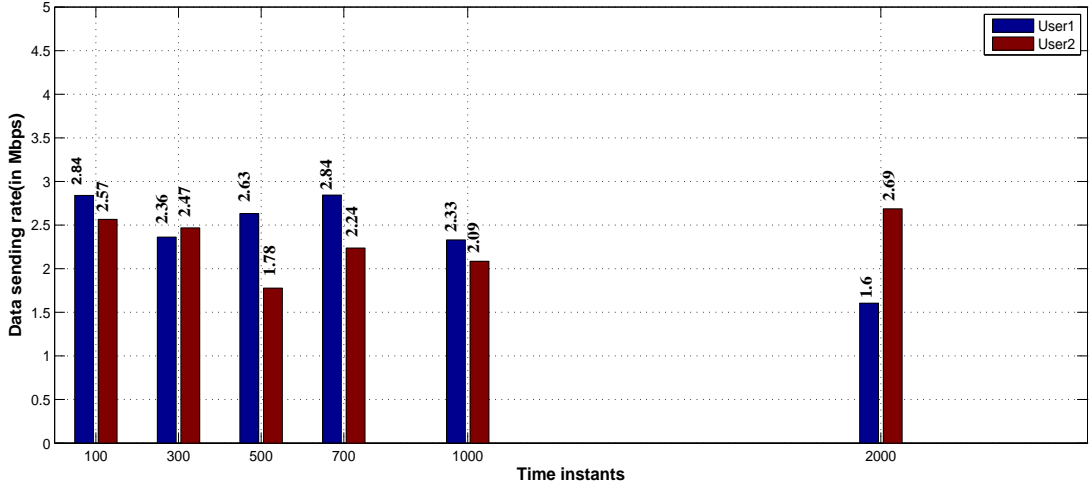}
\caption{Conservation of data sending rates of User1 and User2 at different time instants}
\label{f4}
\end{center} 
\end{figure*}

In the communication networks, many paths are available for sending packets for each user, between the desired source and destination. All routes are assumed as equally weighted hence, users can select any of the available routes for sending the packets.
Each user can send data along one of the shortest paths to the destination with a maximum flow rate of individual links. The data sending rate is reduced as multiple users want to share the common resources. Furthermore, user may not send more data with maximum rate. User's rate depends on two parameters; it's own willingness to pay $  \mathcal{P}_r(t_i) $ and network's feedback $ \psi_r(t_i) = \sum_{e \in r}\psi_e(t_i) $. Using rate control theorem given in Eq. (\eqref{e4}), an optimal data sending rate of each user is obtained (Fig. \ref{f4}). Optimal data rate of two users, User1  and User2 are shown at different time instants. Interval of the two time instants $ \Delta t $ is taken as $ \Delta t = 5 * 10^2 $. Size of the network is changing but number of users are kept same as in initial network. Size of the network increases but optimal rates are decreasing to a stable point. User rates depend on the demand of particular resources coming in the shortest route. If demand is high then, data sending rate will be less.
  
%

Multiple users want to establish connections between a distinct pair of nodes through a shortest possible communication path. More than one shortest routes are possible for making those connection, so data flow rates will be less through the paths having highest betweenness centrality and lowest reputation summation of all the existing nodes. A network is formed with the network network of size $ N = 5 * 10^2 $ with considering $ 20 $ source to destination $ S-D $ pairs along with their betweenness centrality and reputation values and optimal rate ($ x^* $) (Table \ref{tab3}). 

\begin{table*}[!htb]
\caption{Optimal rate $ x_r^* $ of $ 20 $ users through the shortest routes with minimum and maximum $ \theta_{gr}[\sigma(S \rightarrow D)] $ (maximum and minimum), when size of network $ N = 5 * 10^2 $} 
 \center
\begin{tabular}[1]{l| c| c| c| c c| c c} 
\hline\hline 
& Source & Destination & path & $ \theta_{gr}[\sigma(S \rightarrow D)] $ & & $ \theta_{gr}[\sigma(S \rightarrow D)] $ & \\
& & & length &  (Minimum)  &  ($ x_r^* $ )  & (Maximum)  & ( $ x_r^* $)
\\ [0.5ex]
\hline 
User1 & 29	 & 14 & 3 & 0.0156 & 8.9969  & 0.0987 & 3.3189\\
\hline
User2 & 77	& 1	& 2	& 0.0192	 & 8.0567 &  0.0552 & 3.8823 \\
\hline
User3 & 74 & 73 & 4 & 0.0025 & 4.8526 &  0.0509 & 2.2117 \\
\hline
User4 & 72 & 1 & 2 & 0.0183 & 8.7746	& 	0.0544 & 	4.0416\\
\hline
User5 & 58 & 73 & 3 & 0.0118 	& 4.5454 & 0.0518	& 	2.7566\\
\hline
User6 & 34 & 84 & 3 & 0.0115 	& 8.8328	 & 0.0560	 &  4.4339\\
\hline
User7 & 94 & 25 & 3 & 0.0028 & 10.8234 & 0.0508 & 3.1719\\
\hline
User8 & 96 & 36 & 2 & 0.0005	& 4.8525 &  0.0008	 & 3.6785\\
\hline
User9 & 66 & 85 & 4 & 0.0027 & 6.1431 & 	0.0566 & 2.4510\\
\hline
User10 & 20 & 2 & 3 & 0.0023	& 9.2059 &  0.0569 & 4.4499\\
\hline
User11 & 58 & 16 & 3 & 0.0121 & 8.0484 & 0.0554 & 3.6219\\
\hline
User12 & 3 & 52 & 2 & 0.0415	& 8.5054 & 	 0.0457 &	5.2154\\
\hline
User13 & 	14 & 	26 & 	3	& 0.0100 & 9.2094 & 	0.0646 & 	4.6149\\
\hline
User14 & 	95 & 	74 & 	5	& 0.0006 & 5.7869 & 		0.0517 & 	2.7312\\
\hline
User15 & 	70 & 	58 & 	3   & 0.0025 & 6.2374	&  	0.0559 & 	3.5818\\
\hline	
User16 &    31 & 	69 & 	3 & 	0.0024 & 7.0966 & 	 	0.0578 & 	4.9822\\
\hline
User17 & 28 & 	12 & 	3	& 0.0072 &	7.2207 & 	0.0582 & 	2.7948\\
\hline
User18 & 22 & 	7 &	3 &	0.0236	& 6.6195 &  0.0852 & 	2.7941\\
\hline
User19 & 	89	& 100	& 3 & 	0.0030 & 8.6138 & 	 	0.0503 & 	3.7978\\
\hline
User20 & 	57 & 	31 & 3 & 	0.0025 & 8.0771 & 	 	0.0535 & 	4.0460 \\
\hline

\end{tabular}
\label{tab3}
\end{table*}

\begin{figure*}[!htb]
\begin{center}
$\begin{array}{cc}
\includegraphics[width=0.5\linewidth, height=2.8 in]{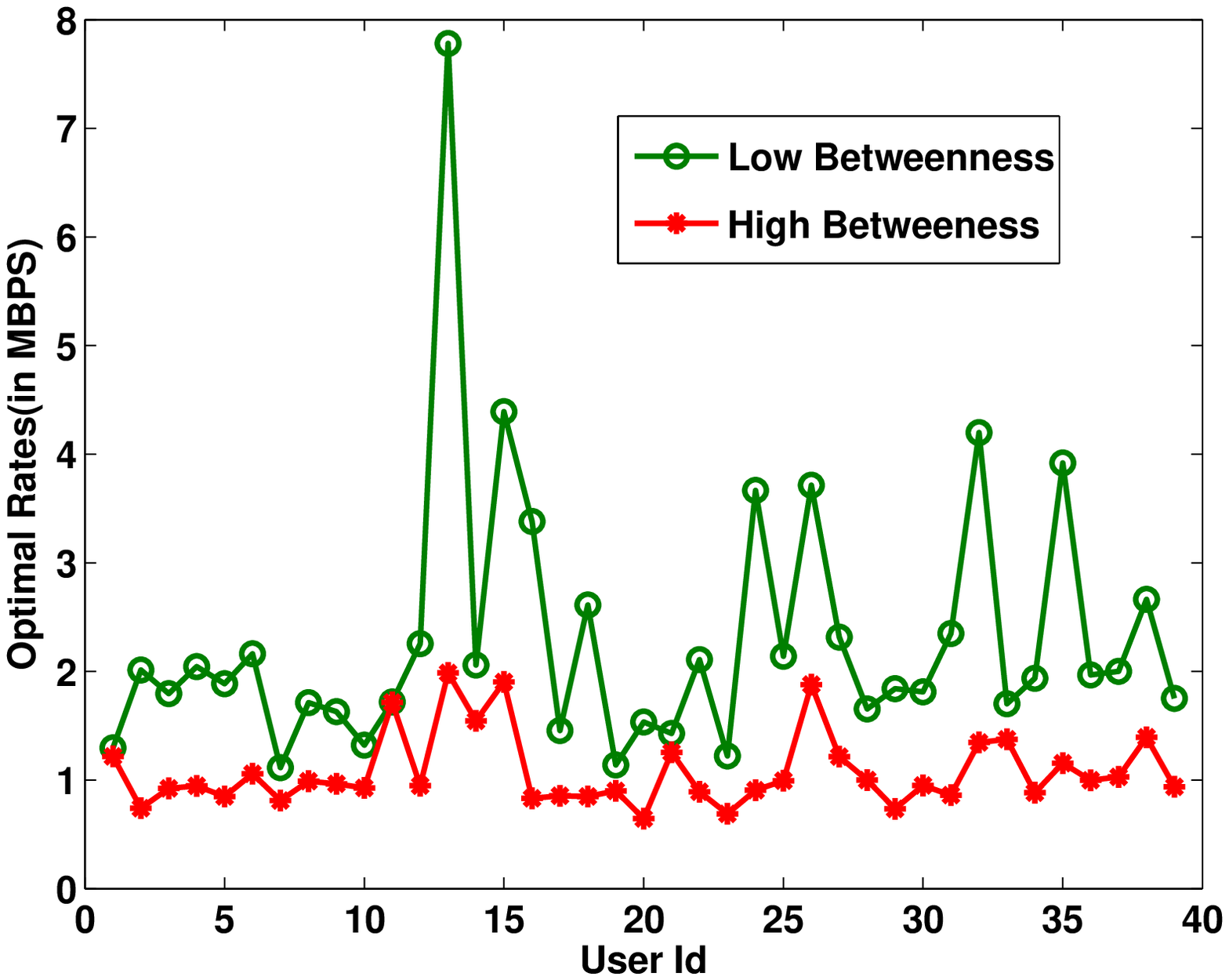} &
\includegraphics[width=0.5\linewidth, height=2.8 in]{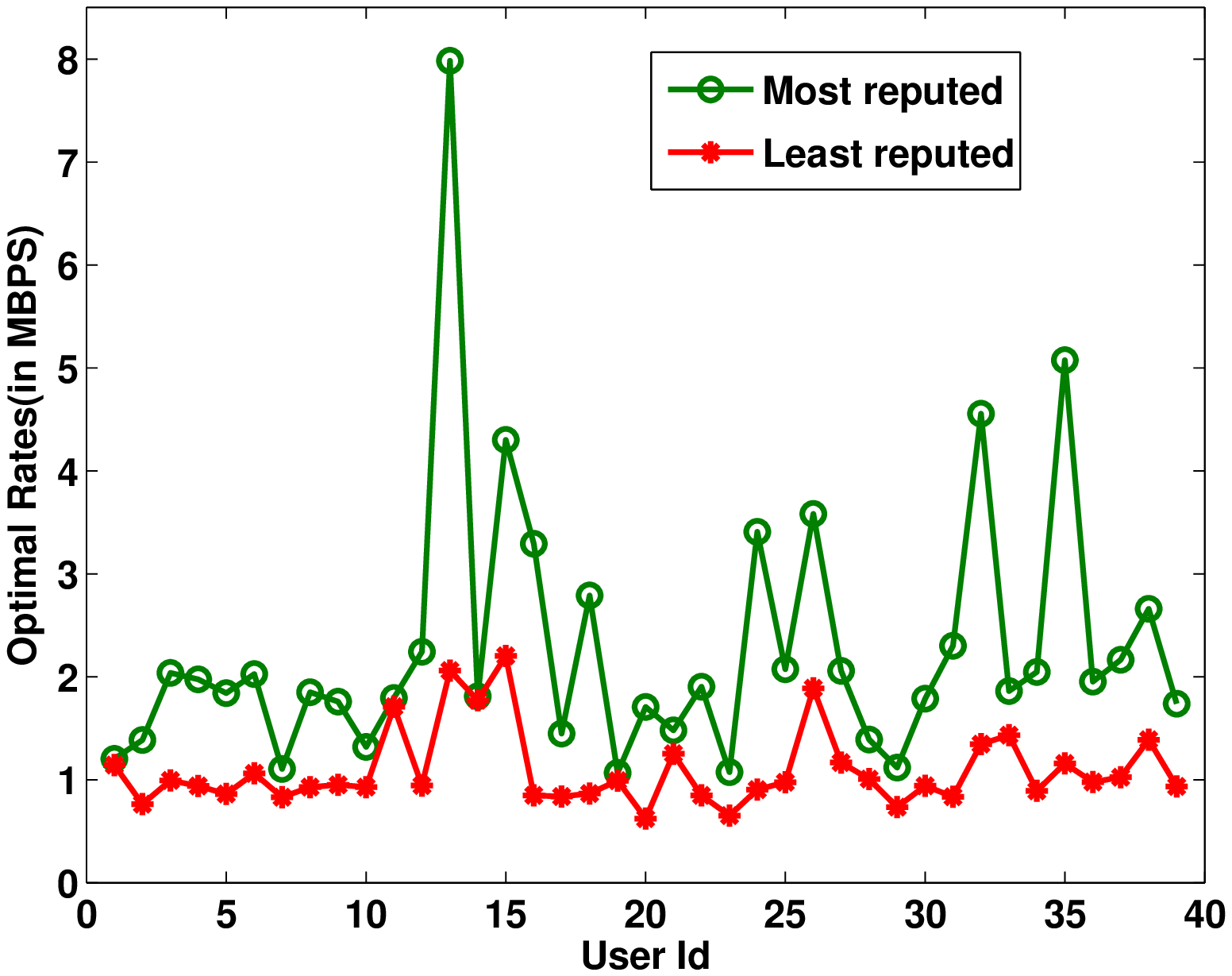} \\
\end{array}$
\caption{User's optimal rates of user paths with \textbf{(a)} highest \& lowest betweenness and \textbf{(b)} reputation (highest \& lowest)}
\label{f5}
\end{center} 
\end{figure*}

%
Now, the number of users are equal to the network size $ N = 5 * 10^2 $. Here, in this network $ 39 $ users' source destination pairs are taken and for each pair the highest and lowest summation of the betweenness centralities for all the available shortest path are calculated. Optimal data sending rates of the $ 39 $ users with highest and lowest betweenness centrality are calculated and the result are shown in Fig. \ref{f5}(a). Apart from betweenness centrality, there is one more parameter named as \textit{reputation} effects optimal data rate $ x^* $ of each user. If path is formed by takings over all sum of most reputed nodes then, rate of the user will be maximum. Results of optimal data rates ($ x^* $) of $ 39 $ users with aggregate sum of most and least reputed nodes are displayed in Fig.\ref{f5}(b).

\begin{figure*}[!htb]
\begin{center}
\includegraphics[width=0.75\linewidth, height=3 in]{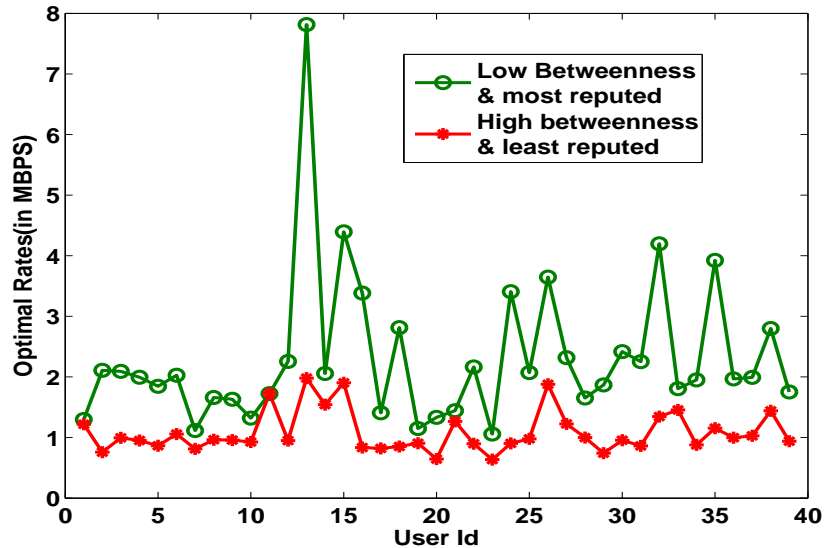}
\caption{User's optimal rates at time $ t = 5* 10^2 $ by considering both betweenness and reputation ( $ \theta_{gr}[\sigma(S \rightarrow D)] $) with maximum \& minimum value}
\label{f6}
\end{center} 
\end{figure*}

Calculation of betweenness centrality and reputation individually is not sufficient for obtaining optimal rates of users. Hence, we have considered both betweenness and reputation of the nodes in the networks. Those paths are selected. A path considered as efficient for routing if contains those nodes, which is least betweenness central as well as reputed.  Simulation results in Fig. \ref{f6} show that the data sending rate of the user path having lowest or highest value for the aggregate sum of betweenness and highest or lowest total sum of reputation, is always maximum or minimum. 

\section{Conclusions and future work}
A framework is designed to represent time varying communication networks (TVCN) where links may appear/disappear randomly with time and nodes are added continuously. A model is proposed to design an optimal network with the consideration of all the aspects of dynamicity (addition, rewiring and removal) of links. At each time step, a new node is added into the network, and a number $ X $  is taken for establishing, rewiring and removal of connections. The values of the fraction $ \beta $ and $ \gamma $ determine the number of added and altered (rewired and removed) links. Various experiments were performed for finding the topological structure for the TVCN with different values of correlation parameter, $ \delta $. It is observed that with increasing value of $ \beta $ in the range $ (0, 1) $ leads to decrement of the correlation parameter $ \delta $ in the interval $ (\infty, 0) $. Furthermore, the scaling exponent increases in the range $ 2 < \alpha \leq 3 $. It is found that the value of power law exponents obtained from the simulation and mean field theoretical analysis of connectivity distribution are proximate to each other. In TVCN, multiple users select their paths such that nodes in the path should have least betweenness centrality and most reputed. Once a path is established, optimal data sending rate of the user is obtained by using rate control theorem proposed by Kelly \cite{kelly2001}. Interval of the two time instants $ (\Delta t) $ is taken as $ \Delta t = 5 * 10^2 $. Number of users are kept same as in initial network despite of change in network size. At the end of $ (t_i+\Delta t)^{th} $ time instant, $ x^* $ of users are evaluated and it is found that size of the network increases but optimal rates are decreasing to a stable point, $ x^* $. At each time step, optimal rate of the users are calculated and it is found that data sending rate of the user's are different with time. We have also studied the case in which number of users increase with network growth. Shortest path with lowest or highest betweenness and highest or lowest reputation are chosen for routing and observed that the value of optimal rate will always be the maximum or minimum.

In this paper, system utility is calculated without consideration of delays in the network. Utility function, itself can be replaced by using game theoretical approach. We have not considered the resilience under cascaded failure and targeted attack in the network.

\bibliographystyle{splncs}
\bibliography{dynamic_net}

\end{document}